\documentclass[12pt]{article}
\usepackage{latexsym,amssymb,graphics}
\usepackage{latexsym,amsfonts,amssymb}

\def\nn{\nonumber}

\author{Renat Zhdanov\thanks{E-mail:\ renat.zhdanov@bio-key.com}\\
BIO-key International, Eagan, MN, USA}
\title{Nonlocal symmetries of evolution equations.}
\date{}

\begin{document}
\maketitle

\begin{abstract}
We suggest the method for group classification of evolution
equations admitting nonlocal symmetries which are associated with a
given evolution equation possessing nontrivial Lie symmetry. We
apply this method to second-order evolution equations in one spatial
variable invariant under Lie algebras of the dimension up to three.
As a result, we construct the broad families of new nonlinear
evolution equations possessing nonlocal symmetries which in
principle cannot be obtained within the classical Lie approach.
\end{abstract}

\section{Introduction}

Classical Lie symmetries has become an inseparable part of the
modern mathematical physics toolkit for analysis of linear and
nonlinear differential equations \cite{lie1}-\cite{zhd0}. However,
with all its power, versatility and universality the Lie symmetry
approach has its limits, which prompted numerous researchers to look
for possible generalizations of the concept of classical symmetry in
order to be able to handle differential equations that do not
possess Lie symmetries.

One of the promising generalizations of the concept of the classical
symmetry is allowing for a generalized symmetry to depend
on integrals of dependent variables, in contrast to Lie symmetry
which can only involve independent, dependent variables and their
derivatives. In this way nonlocal symmetries have been introduced
into the modern mathematical physics.

There was significant progress in studying nonlocal symmetries of
linear differential equations (see, e.g., \cite{fus1} and the
references therein). Much less is known about nonlocal symmetries of
nonlinear differential equations.

A possible way to approach the problem of constructing nonlocal
symmetries of partial differential equations is utilization of the
concept of quasi-local symmetry introduced in \cite{ibr0,puk}.
Quasi-local symmetry is the one obtained from a classical Lie
symmetry through nonlocal transformation. Another approach is based
on the concept of potential symmetry \cite{blu1}-\cite{blu3}.
However, as we established recently, potential symmetries of
evolution type equations are quasi-local ones in a sense that there
is a nonlocal transformation reducing those to point or contact
symmetries \cite{zhd1}.

The principal motivation for this paper was a need for a more
systematic approach to classification of partial differential
equations admitting quasi-local symmetries. It is natural to expect
that such an approach should rely on Lie symmetries of equations
under study. We demonstrate in the paper that it is indeed possible
to develop the regular and systematic classification scheme using
the results of group classification of nonlinear second-order
equations \cite{zhd2,zhd3} and the ideas suggested in
\cite{zhd1,sok1}.

We utilize the results of group classification of the general
second-order evolution equation in one spatial variable
\begin{equation}
\label{0.1}
u_t=f(t,x,u,u_x,u_{xx})
\end{equation}
to construct nonlocal symmetries of the associated evolution
equations. In a what follows, we denote the class of partial
differential equations (\ref{0.1}) as ${\mathbb E}$.

The paper is organized as follows. In Section 2 we give the
necessary definitions and present the detailed description of our
approach to classifying nonlocal symmetries of (\ref{0.1}). Section
3 is devoted to application of the approach to equations from
${\mathbb E}$ invariant under the Lie algebras of the dimension
$s\le 3$. The last section contains brief discussion of the obtained
results.

\section{Description of the method}

We begin by introducing the differential transformation which plays
a key role in our subsequent considerations
\begin{equation}
\label{1.1}
{\mathfrak P}(t,x,u)=(\bar t,\bar x,\bar u) \equiv (t,x,u_x).
\end{equation}
Note that the inverse of differential transformation (\ref{1.1})
reads as
\begin{equation}
\label{1.2}
t = \bar t,\quad x = \bar x,\quad u = \partial_{\bar x}^{-1}\bar u.
\end{equation}
Hereafter $\partial_x^{-1}$ is the inverse of the differentiation
operator $\partial_x$, i.e., $\partial_x\,\partial_x^{-1} \equiv
\partial_x^{-1}\,\partial_x \equiv 1$.

Consider a subclass of evolution equations of the form (\ref{0.1})
\begin{equation}
\label{1.3}
u_t=f(t,x,u_x,u_{xx})
\end{equation}
admitting the group of translations by $u$.

Differentiating (\ref{1.3}) with respect to $x$ and applying
$\mathfrak P$ yields
\begin{equation}
\label{1.4}
\bar u_{\bar t} = \partial_{\bar x} f(\bar t,\bar x,\bar u,\bar
u_{\bar x}).
\end{equation}
Consequently, ${\mathfrak P}$ maps a subclass of equations
(\ref{1.3}) from ${\mathbb E}$ into another subclass of equations
(\ref{1.4}) from ${\mathbb E}$. Remarkably, this very transformation
is responsible for all potential symmetries of equations (\ref{0.1})
\cite{zhd1}.

Indeed, suppose that (\ref{1.3}) admits Lie transformation group
\begin{equation}
\label{1.5}
t' = T(t,\theta),\quad x' = X(t,x,u,\theta),\quad u' =
U(t,x,u,\theta),
\end{equation}
where $\theta$ is the group parameter. Note that (\ref{1.5}) is the
most general one-parameter Lie group that can be admitted by an
equation from the class ${\mathbb E}$ (see, e.g., \cite{zhd4}).
Computing the first prolongation of formulas (\ref{1.5}) we derive
the transformation rule for the first derivative of $u$
$$
\frac{\partial u'}{\partial x'}=\frac{U_uu_x+U_x}{X_uu_x+X_x}.
$$
Now the image of transformation group (\ref{1.4}) under the action
of ${\mathfrak P}$ takes the form
\begin{equation}
\label{1.6}
\bar t' = T(\bar t,\theta),\quad \bar x' = X(\bar t,\bar
x,u,\theta)),\quad \bar u'=\frac{U_{u}\bar u+U_{\bar x}}{X_{u}\bar
u+X_{\bar x}}
\end{equation}
with $u=\partial_{\bar x}^{-1}\bar u$ and $U=U(\bar t,\bar
x,u,\theta)$. Consequently, if the right-hand sides of (\ref{1.6})
contain nonlocal variable $u$, then (\ref{1.6}) is nonlocal symmetry
of (\ref{1.3}). According to \cite{zhd1}, transformations
(\ref{1.6}) include the variable $u$ if and only if $X_u\not=0$ or
$U_{\bar xu}^2+U_{uu}^2\not= 0$.

Summing up we conclude that provided symmetry group (\ref{1.5}) of
equation (\ref{1.3}) obeys one of the constraints
\begin{equation}
\label{1.7.1}
X_u\not=0\ {\rm or}\ U_{xu}^2 + U_{uu}^2\not= 0,
\end{equation}
then ${\mathfrak P}$ maps this symmetry into nonlocal symmetry
(\ref{1.6}) of equation (\ref{1.4}). This nonlocal symmetry is
called the potential symmetry of (\ref{1.4}).

Constraints (\ref{1.7.1}) can be written in terms of the
coefficients of the infinitesimal operator of group (\ref{1.5})
\begin{equation}
\label{1.8}
S=\tau(t)\partial_t + \xi(t,x,u)\partial_x + \eta(t,x,u)\partial_u.
\end{equation}
If
\begin{equation}
\label{1.7.2}
\xi_u\not=0\ {\rm or}\ \eta_{xu}^2 + \eta_{uu}^2\not= 0,
\end{equation}
then symmetry $S$ is mapped by ${\mathfrak P}$ into nonlocal
symmetry \cite{zhd1}.

Despite the fact that equation (\ref{1.3}) is rather particular case
of the general evolution equation (\ref{0.1}), the scheme outlined
above provides a very general framework for group classification of
nonlocal symmetries.

Let the symbol ${\mathcal S}$ stand for the infinite-dimensional Lie
algebra of differential operators (\ref{1.8}) and the symbol
${\mathcal I}$ stand for the infinite-dimensional Lie algebra
spanned by the operators of the form
\begin{equation}
\label{1.9}
Q=\xi(t,x,u)\partial_x + \eta(t,x,u)\partial_u.
\end{equation}
As direct computations shows, $[{\mathcal I},{\mathcal I}]\subset
{\mathcal I}$ and $[{\mathcal I},{\mathcal S}]\subset {\mathcal I}$,
which means that ${\mathcal I}$ is the ideal in the Lie algebra
${\mathcal S}$.

If an equation of the form (\ref{0.1}) possesses Lie symmetry from
${\mathcal I}$, then it can always be reduced to the form
(\ref{1.3}) by a suitable local transformation of variables
\cite{ovs1}. Consequently, the procedure for classification of
nonlocal symmetries described above is applicable to any equation
from the class ${\mathbb E}$ admitting at least one symmetry from
${\mathcal I}$.

In the paper \cite{zhd1} we suggest a generalization of the
potential symmetry approach which allows for the use of more general
differential transformations to construct nonlocal symmetries. Our
generalization is based on the idea put forward by Sokolov
\cite{sok1}. He suggests that a differential transformation is to be
looked for in the form
\begin{equation}
\label{1.10}
{\mathfrak D}(t,x,u)=(\bar t,\bar x,\bar u) \equiv
\Bigl(t,\omega_1(t,x,u,u_x,u_{xx},\ldots),
\omega_2 (t,x,u,u_x,u_{xx},\ldots)\Bigr),
\end{equation}
$\omega_1$ and $\omega_2$ being differential invariants of the
symmetry group of the evolution equation under study. For the case
when ${\mathfrak D} = {\mathfrak P}$, we have $\omega_1=x$ and
$\omega_2=u_x$, $x$ and $u_x$ being the simplest differential
invariants of the group of displacements by $u$ generated by the
infinitesimal operator $\partial_u$. Evidently, evolution equation
(\ref{1.3}) admits symmetry $\partial_u$. Sokolov proves in
\cite{sok1} that for any contact symmetry of evolution equation of
the form (\ref{0.1}) preserving the temporal variable $t$ the
corresponding differential transformation maps equation in question
into ${\mathbb E}$.

Now we are ready to formulate the group approach to classifying
nonlocal symmetries of equations from the class ${\mathbb E}$.

Let equation (\ref{0.1}) be invariant under the $r$-dimensional Lie
algebra ${\mathcal L}_r=\langle e_1, \ldots, e_r\rangle$. Next,
suppose that ${\mathcal L}_r$ contains a subalgebra ${\mathcal I}_s$
$=$ $\langle e_1, \ldots, e_s\rangle$,\ $s<r$ such that ${\mathcal
I}_s$ $\subset$ ${\mathcal I}$. Saying it another way, the basis
elements of the algebra ${\mathcal I}_s$ have the form (\ref{1.9}).
Let $t, \omega_1, \omega_2$ be functionally independent differential
invariants of ${\mathcal I}_s$.

Making differential transformation (\ref{1.10}) with so chosen
$\omega_1, \omega_2$ we map the invariant equation (\ref{0.1}) into
another equation from ${\mathbb E}$. If the transformation group
generated by the operators $e_{s+1},\ldots, e_r$ obeys one of the
inequalities (\ref{1.10}), then its image under the differential
transformation ${\mathfrak D}$ is a nonlocal symmetry of the
transformed equation.

Direct application of the transformation of the form (\ref{0.1}),
especially for the case when $\omega_1, \omega_2$ depend on higher
derivatives, may become quite challenging computational problem by
itself. That is why we suggest a modification of the approach above,
which is to represent ${\mathfrak D}$ as a superposition of
elementary differential transformations. These elementary
transformations are obtained as a superposition of a local
transformation reducing the individual basis operators of the Lie
algebra ${\mathcal I}_s$ to the form $\partial_u$ and of the
differential transformation ${\mathfrak P}$.

We begin with the local change of variables, ${\mathfrak T}_1$,
reducing the first basis operator $e_1$ to the form $\partial_u$ and
after that we apply the transformation ${\mathfrak P}$. Next, we
recompute the remaining basis operators in the new variables
$t,x,u_x$.

To construct the image of infinitesimal operator (\ref{1.9}) under
the transformation ${\mathfrak P}$ we compute the coefficient of
$\partial_{u_x}$ in its first prolongation
$$
\zeta(t,x,u,u_x) = \eta_x + u_x(\eta_u - \xi_x) -u_x^2\xi_{uu}.
$$
Next, we replace $u$ with $\partial_x^{-1}u$ and $u_x$ with $u$ so
that the transformed infinitesimal operator, ${\bar Q}$, takes the
form
\begin{equation}
\label{1.11}
\bar Q = \xi(t,x,u)\partial_x +
\zeta(t,x,\partial_x^{-1}u,u)\partial_u.
\end{equation}
If there is a basis element, $e_i$, which does not depend on the
nonlocal variable $\partial_x^{-1}u$, then we can repeat the
procedure and reduce $e_i$ to the form $\partial_u$ and then apply
${\mathfrak P}$ and so on.

The above described procedure has two possible outcomes. The first
one is when we are able to perform all $s$ steps and get an
evolution equation which is the image of the initial ${\mathcal
I}_s$-invariant equation under transformation (\ref{1.10}). The
resulting differential transformation takes the form
\begin{equation}
\label{1.12}
{\mathfrak T} = \prod_{i=s}^1\,({\mathfrak P} \circ
{\mathfrak T}_i) \equiv {\mathfrak P} \circ {\mathfrak T}_s \circ
{\mathfrak P} \circ {\mathfrak T}_{s-1} \circ \cdots \circ
{\mathfrak P} \circ {\mathfrak T}_1.
\end{equation}
Applying ${\mathfrak T}$ to (\ref{0.1}) yields the sequence of $ s$
evolution equations from the class ${\mathbb E}$ together with the
sequence of their symmetries, some of which might be nonlocal.

The second outcome is when at some step $p < s$ all the remaining
basis operators contain the nonlocal variable $\partial_x^{-1}u$ and
we cannot proceed any further. In this case we obtain the sequence
of transformations
\begin{equation}
\label{1.13}
{\mathfrak T} = \prod_{i=p}^1\,({\mathfrak P} \circ
{\mathfrak T}_i) \equiv {\mathfrak P} \circ {\mathfrak T}_p \circ
{\mathfrak P} \circ {\mathfrak T}_{p-1} \circ \cdots \circ
{\mathfrak P} \circ {\mathfrak T}_1
\end{equation}
that maps the initial ${\mathcal I}_s$-invariant evolution equation
into equation, which admits at least $s-p$ nonlocal symmetries of
the form (\ref{1.11}).

We summarize the above speculations in the form of the algorithm of
classification of evolution equations possessing nonlocal
symmetries.

The starting point of our classification algorithm is evolution
equation (\ref{0.1}) invariant under $r$-dimensional Lie algebra
${\mathcal L}_r \subset {\mathcal S}$. Suppose also that the algebra
${\mathcal L}_r$ contains a subalgebra ${\mathcal I}_s \subset
{\mathcal I}$ of the dimension $1\le s \le r$. Now to classify
evolution equations admitting nonlocal symmetries one needs to
\begin{itemize}

\item{classify ${\mathcal I}_s$-inequivalent subalgebras ${\mathcal
J}^1, \ldots, {\mathcal J}^N$ of the algebra ${\mathcal I}_s$,}

\item{compute the sequence of transformations (\ref{1.12}) mapping
the initial invariant equation into ${\mathbb E}$ for each
subalgebra ${\mathcal J}^i,\ i=1, \ldots, N$ and generate the
corresponding sequence of evolution equations and their invariance
algebras,}

\item{verify for each invariance algebra whether its basis elements
satisfy one of the inequalities (\ref{1.7.2}). If this the case,
then the remaining elements of the sequence of evolution equations
possess nonlocal symmetry.}

\end{itemize}

Upon completing the three steps of the algorithm above ones obtains
the set of evolution equations (\ref{0.1}) that admit nonlocal
symmetries.

\section{Applications}

In \cite{zhd2,zhd3,zhd4} we obtain exhaustive description of
inequivalent Lie algebras, which are symmetry algebras of equations
of the form (\ref{0.1}). This fact facilitates application of the
algorithm presented in the previous section in its full generality.
However, in this paper we restrict our considerations to subalgebras
of the symmetry algebra of (\ref{0.1}) of the dimension $s\le 3$.

The list of inequivalent realizations of one-dimensional symmetry
algebras of equation (\ref{0.1}) consists of one representative
$$
{\mathcal A}_1=\langle \partial_u \rangle:\quad u_t =
f(t,x,u_x,u_{xx}).
$$
This case has already been considered in the previous section, the
algebra ${\mathcal A}_1$ giving rise to the transformation
${\mathfrak P}$.

According to \cite{zhd4} the class of operators, ${\mathcal I}$,
contains four inequivalent realizations of two-dimensional symmetry
algebras of (\ref{0.1}). These realizations are listed below
together with the corresponding invariant equations
\begin{eqnarray*}
&&{\mathcal A}_2^1 = \langle \partial_x, \partial_u \rangle:\quad
u_t = f(t,u_x,u_{xx}),\\
&&{\mathcal A}_2^2 = \langle x \partial_u, \partial_u \rangle:\quad
u_t = f(t,x,u_{xx}),\\
&&{\mathcal A}^3_2 = \langle -x \partial_x -u \partial_u, \partial_u
\rangle:\quad u_t = xf(t,u_x,x u_{xx}),\\
&&{\mathcal A}^4_2 = \langle -u\partial_u, \partial_u \rangle:\quad
u_t = u_xf(t,x,u^{-1}_xu_{xx}).
\end{eqnarray*}

We start by analyzing differential transformations associated with
the algebra ${\mathcal A}_2^1$. Applying the transformation
${\mathfrak P}$ to the equation $u_t = f(t,u_x,u_{xx})$ we get
$$
u_t=f_u + u_{xx}f_{u_x},
$$
where $f=f(t,u,u_x)$. Note that we dropped the bars.

The second basis operator of ${\mathcal A}_2^1$ reads now as
$\partial_x$. To reduce it to the the form $\partial_u$ we apply the
hodograph transformation
\begin{equation}
\label{3.1}
{\mathfrak T}_1(t,x,u) = (\bar t,\bar x,\bar u)\equiv (t,u,x).
\end{equation}
The transformed equations has the form
$$
u_t = -f_{z_1} + \frac{u_{xx}}{u_x}f_{z_2},
$$
where $f = f(t,z_1,z_2) \equiv f(t,u,-u_x^{-1})$. Applying the
transformation ${\mathfrak P}$, yields the final form of the
transformed evolution equation
$$
u_t = \frac{\partial}{\partial x}\Bigl(-f_{z_1} +
\frac{u_x}{u}f_{z_2}\Bigr).
$$
Here $f = f(t,z_1,z_2) \equiv f(t,u,-u^{-1})$.

Turn now to the algebra ${\mathcal A}_2^2$. Applying the
transformation ${\mathfrak P}$ to the invariant equation $u_t =
f(t,x,u_{xx})$ yields
$$
u_t=f_x + u_{xx}f_{u_x}
$$
with $f=f(t,x,u_x)$. As usual we drop the bars. The second basis
element of ${\mathcal A}_2^2$ takes the form $\partial_u$ and we can
directly apply the transformation ${\mathfrak P}$ thus getting
$$
u_t=f_{xx} + 2u_xf_{xu} + u_x^2f_{uu} + f_uu_{xx}.
$$

Consider next the algebra ${\mathcal A}_2^3$. Applying the
transformation ${\mathfrak P}$ to the equation $u_t = xf(t,u_x,x
u_{xx})$ we have
\begin{equation}
\label{3.2}
u_t=f + xu_xf_{z_1} + x(u_x + xu_{xx})f_{z_2}
\end{equation}
with $f = f(t,z_1,z_2) \equiv f(t,u,xu_x)$. The second basis
operator of the algebra ${\mathcal A}_2^3$ reads now as
$x\partial_x$. Applying the hodograph type local transformation
\begin{equation}
\label{3.3}
{\mathfrak T}_2(t,x,u) = (\bar t,\bar x,\bar u)\equiv (t,u,\ln x)
\end{equation}
we reduce it to the form $\partial_u$. Transforming equation
(\ref{3.2}) accordingly we get
$$
u_t = -xf_{z_1} + (x^2u_x^{-2}u_{xx} + xu_x^{-1} - x)f_{z_2},
$$
where $f = f(t,z_1,z_2) \equiv f(t,x,u_x^{-1})$. Now we can apply
the transformation ${\mathfrak P}$ thus getting
$$
u_t = \frac{\partial}{\partial x}\Bigl(-xf_{z_1} +
(x^2u^{-2}u_x + xu^{-1} - x)f_{z_2}\Bigr).
$$
with $f = f(t,z_1,z_2) \equiv f(t,x,u^{-1})$.

Finally, we turn to the algebra ${\mathcal A}_2^4$. Applying the
transformation ${\mathfrak P}$ to the invariant equation $u_t =
u_xf(t,x,u^{-1}_xu_{xx})$ yields
\begin{equation}
\label{3.4}
u_t = u_xf + uf_{z_1} + (u_{xx}-u^{-1}u_x^2)f_{z_2},
\end{equation}
where $f = f(t,z_1,z_2) \equiv f(t,x,u^{-1}u_x)$. The second basis
element of the algebra ${\mathcal A}_2^4$ is still of the form
$u\partial_u$, so we need to reduce it to the form $\partial_u$. To
this end, we perform the local transformation
\begin{equation}
\label{3.5}
{\mathfrak T}_3(t,x,u) = (\bar t,\bar x,\bar u)\equiv (t,x,\ln u).
\end{equation}
As a result, (\ref{3.4}) takes the form
$$
u_t = u_xf + f_{z_1} + u_{xx}f_{z_2}
$$
with $f = f(t,z_1,z_2) \equiv f(t,x,u_x)$. Applying the
transformation ${\mathfrak P}$ finally yields
$$
u_t = u_xf + uf_x + uu_xf_u + f_{xx} + 2u_xf_{xu} + u_x^2f_{uu}
+u_{xx}f_u,
$$
where $f=f(t,x,u)$.

Summing up we present the four sequences of differential
transformations that map four subclasses of the class of
second-order evolution equations ${\mathbb E}$ into ${\mathbb E}$.
\begin{itemize}
\item Differential transformation ${\mathfrak P} \circ {\mathfrak
T}_1 \circ {\mathfrak P}$ maps equation
$$
u_t = f(t,u_x,u_{xx})
$$
into
$$
u_t = \frac{\partial}{\partial x}\Bigl(-f_{z_1} +
\frac{u_x}{u}f_{z_2}\Bigr)
$$
with $f = f(t,z_1,z_2) \equiv f(t,u,-1/u)$.
\item Differential transformation ${\mathfrak P} \circ {\mathfrak
P}$ maps equation
$$
u_t = f(t,x,u_{xx})
$$
into
$$
u_t=f_{xx} + 2u_xf_{xu} + u_x^2f_{uu} + f_uu_{xx}
$$
with $f = f(t,x,u)$.
\item Differential transformation ${\mathfrak P} \circ {\mathfrak
T}_2 \circ {\mathfrak P}$ maps equation
$$
u_t = xf(t,u_x,x u_{xx})
$$
into
$$
u_t = \frac{\partial}{\partial x}\Bigl(-xf_{z_1} +
(x^2u^{-2}u_x + xu^{-1} - x)f_{z_2}\Bigr)
$$
with $f = f(t,z_1,z_2) = f(t,x,u^{-1})$.
\item Differential transformation ${\mathfrak P} \circ {\mathfrak
T}_3 \circ {\mathfrak P}$ maps equation
$$
u_t = u_xf(t,x,u^{-1}_xu_{xx})
$$
into
$$
u_t = u_xf + uf_x + uu_xf_u + f_{xx} + 2u_xf_{xu} + u_x^2f_{uu}
+u_{xx}f_u,\quad f=f(t,x,u).
$$
\end{itemize}
In the above formulas the transformations ${\mathfrak T}_1,
{\mathfrak T}_2, {\mathfrak T}_3$ are given by (\ref{3.1}),
(\ref{3.2}) and (\ref{3.5}), respectively.

The class of first-order differential operators ${\mathcal I}$
contains twenty two inequivalent realizations of three-dimensional
symmetry algebras of (\ref{0.1}) \cite{zhd4}. These realizations are
listed below together with the corresponding invariant equations.
\begin{eqnarray*}
&&{\mathcal A}^1_3 = \langle -x\partial_x-u\partial_u, \partial_u,
xt\partial_x \rangle:\quad u_t = -xt^{-1}u_x\ln u_x + xu_xf(t,x
u^{-1}_x u_{xx}),\\
&&{\mathcal A}^2_3 = \langle -x\partial_x-u\partial_u, \partial_u,
x\partial_x\rangle:\quad u_t = xu_xf(t,xu^{-1}_xu_{xx}),\\
&&{\mathcal A}^3_3 = \langle -x\partial_x-u\partial_u,
\partial_u, x\partial_u \rangle:\quad u_t = xf(t,xu_{xx}),\\
&&{\mathcal A}^4_3 = \langle -u\partial_u, \partial_u,
\partial_x \rangle:\quad u_t = u_xf(t,u^{-1}_xu_{xx}),\\
&&{\mathcal A}^5_3 = \langle \partial_u, \partial_x,
x\partial_u+t\partial_x \rangle:\quad u_t = -\frac{1}{2}u^2_x +
f(t,u_{xx}),\\
&&{\mathcal A}^6_3 = \langle \partial_u, \partial_x, x\partial_u
\rangle:\quad u_t = f(t, u_{xx}),\\
&&{\mathcal A}^7_3 = \langle x\partial_u, \partial_u,
x^2\partial_x+xu\partial_u \rangle:\quad u_t = xf(t,x^3 u_{xx}),\\
&&{\mathcal A}^8_3 = \langle \partial_x, \partial_u,
(x+u)\partial_x+u\partial_u \rangle:\quad u_t =
u_x\exp(u^{-1}_x)f\Bigl(t,u^{-3}_xu_{xx}\exp(u^{-1}_x)\Bigr),\\
&&{\mathcal A}^9_3 = \langle x\partial_u, \partial_u,
x^2\partial_x+u(1+x)\partial_u \rangle:\quad u_t =
x\exp(-x^{-1})f\Bigl(t,xu_{xx}\exp(x^{-1})\Bigr),\\
&&{\mathcal A}^{10}_3 = \langle \partial_x, \partial_u,
x\partial_x+u\partial_u \rangle:\quad u_t = u^{-1}_{xx}f(t,u_x),\\
&&{\mathcal A}^{11}_3 = \langle \partial_x, \partial_u,
x\partial_x-u\partial_u \rangle:\quad u_t =
u_x^{1/2}f(t,u_x^{-3/2}u_{xx}),\\
&&{\mathcal A}^{12}_3 = \langle x\partial_u, \partial_u,
-2x\partial_x-u\partial_u \rangle:\quad u_t =
x^{1/2}f(t,x^{3/2}u_{xx}),\\
&&{\mathcal A}^{13}_3 = \langle \partial_x, \partial_u,
x\partial_x+qu\partial_u \rangle:\quad u_t =
u_x^{q/(q-1)}f(t,u_x^{(2-q)/(q-1)}u_{xx}),\ 0<|q|<1,\\
&&{\mathcal A}^{14}_3 = \langle x\partial_u, \partial_u,
x(q-1)\partial_x+qu\partial_u \rangle:\quad u_t =
x^{q/(q-1)}f(t,x^{(2-q)/(1-q)}u_{xx}),\ q\not=0,\pm 1,\\
&&{\mathcal A}^{15}_3 = \langle \partial_x, \partial_u,
u\partial_x-x\partial_u \rangle:\quad u_t =
(1+u^2_x)^{1/2}f\Bigl(t,(1+u^2_x)^{-3/2}u_{xx}\Bigr),\\
&&{\mathcal A}^{16}_3 = \langle x \partial_u, \partial_u,
(1+x^2)\partial_x+xu\partial_u \rangle:\quad u_t =
(1+x^2)^{1/2}f\Bigl(t,(1+x^2)^{3/2}u_{xx}\Bigr),\\
&&{\mathcal A}^{17}_3 = \langle \partial_x, \partial_u,
(qx+u)\partial_x+(-x+qu) \partial_u \rangle:\quad u_t =
(1+u^2_x)^{1/2}\exp(-q\arctan u_x)\\
&&\quad\times f\Bigl(t,(1+u^2_x)^{-3/2}\exp(-q\arctan
u_x)u_{xx}\Bigr),\ q>0,\\
&&{\mathcal A}^{18}_3 = \langle x \partial_u, \partial_u,
(1+x^2)\partial_x+(x+q)u\partial_u \rangle:\quad u_t =
(1+x^2)^{1/2}\exp(\arctan x)\\
&&\quad\times f\Bigl(t,(1+x^2)^{3/2}\exp(-q\arctan x)u_{xx}\Bigr),\
q\not =0,\\
&&{\mathcal A}^{19}_3 = \langle \partial_u, \cos u\partial_x + \tan
x\sin u\partial_u, -\sin u\partial_x + \tan x\cos u\partial_u
\rangle:\quad\\
&&\quad u_t = (\sec^2 x + u^2_x)^{1/2}f(t,(u_{xx}\cos x - (2 +
u^2_x\cos^2x)u_x\sin x)(1+u^2_x \cos^2 x)^{-3/2}),\\
&&{\mathcal A}^{20}_3 = \langle 2u\partial_u - x\partial_x, -
u^2\partial_u + xu\partial_x, \partial_u \rangle:\quad
u_t = xu_xf(t,x^{-5}u^{-3}_xu_{xx} + 2x^{-6}u^{-2}_x),\\
&&{\mathcal A}^{21}_3 = \langle 2u\partial_u - x\partial_x, (x^{-4}
-u^2)\partial_u + xu \partial_x, \partial_u \rangle:\quad u_t =
x^{-2}(4 + x^6 u^2_x)^{1/2}\\
&&\quad\times f\Bigl(t,(4 + x^6u^2_x)^{-3/2}
(x^4u_{xx} + 5x^3u_x + \frac{1}{2}x^9u^3_x)\Bigr),\\
&&{\mathcal A}^{22}_3 = \langle 2u\partial_u - x\partial_x, -(x^{-4}
+ u^2)\partial_u + xu\partial_x, \partial_u \rangle:\quad u_t =
x^{-2}(x^6u^2_x - 4)^{1/2}\\
&&\quad\times f\Bigl(t,(x^6u^2_x - 4)^{-3/2}(x^4u_{xx} + 5x^3u_x -
\frac{1}{2}x^9u^3_x)\Bigr).
\end{eqnarray*}
Analysis of the sequences of differential transformations associated
with the above algebras is similar to the case of two-dimensional
algebras. We skip the derivation details and give the final result,
the differential transformations (\ref{1.12}), (\ref{1.13}).

For the algebras ${\mathcal A}^1_3$ - ${\mathcal A}^6_3$, ${\mathcal
A}^{10}_3$ - ${\mathcal A}^{14}_3$ The sequence of transformations
${\mathfrak T}$ reads as
\begin{equation}
\label{3.6}
{\mathfrak T} = {\mathfrak P} \circ {\mathfrak T}_2 \circ
{\mathfrak P} \circ {\mathfrak T}_1 \circ {\mathfrak P}.
\end{equation}
where
\begin{eqnarray*}
&&{\mathcal A}^1_3:\quad {\mathfrak T}_1(t,x,u) = (t,u,-\ln{x}),\quad
{\mathfrak T}_2(t,x,u) = (t,xu,t^{-1}\ln{x}).\\
&&{\mathcal A}^2_3:\quad {\mathfrak T}_1(t,x,u) = (t,u,-\ln{x}),\quad
{\mathfrak T}_2(t,x,u) = (t,xu,\ln{x}).\\
&&{\mathcal A}^3_3:\quad {\mathfrak T}_1(t,x,u) = (t,x,u),\quad
{\mathfrak T}_2(t,x,u) = (t,xu,\ln{x}).\\
&&{\mathcal A}^4_3:\quad {\mathfrak T}_1(t,x,u) = (t,x,-\ln{u}),\quad
{\mathfrak T}_2(t,x,u) = (t,xu,\ln{x}).\\
&&{\mathcal A}^5_3:\quad {\mathfrak T}_1(t,u,x) = (t,u,x),\quad
{\mathfrak T}_2(t,x,u) = (t,u,x).\\
&&{\mathcal A}^6_3:\quad {\mathfrak T}_1(t,x,u) = (t,x,u),\quad
{\mathfrak T}_2(t,x,u) = (t,u,x).\\
&&{\mathcal A}^{10}_3:\quad {\mathfrak T}_1(t,x,u) = (t,u,x),\quad
{\mathfrak T}_2(t,x,u) = (t,x,\ln{u}).\\
&&{\mathcal A}^{11}_3:\quad {\mathfrak T}_1(t,x,u) = (t,u,x),\quad
{\mathfrak T}_2(t,x,u) = (t,x^3u^2,-\frac{1}{3}\ln{u}).\\
&&{\mathcal A}^{12}_3:\quad {\mathfrak T}_1(t,x,u) = (t,u,x),\quad
{\mathfrak T}_2(t,x,u) = (t,x^3u^2,-\frac{1}{3}\ln{u}).\\
&&{\mathcal A}^{13}_3:\quad {\mathfrak T}_1(t,x,u) =
(t,u,x),\quad {\mathfrak T}_2(t,x,u) = (t,x^{q-2}u^{q-1},
\frac{1}{q-2}\ln{u}),\quad q \ne 2,\\
&&\phantom{{\mathcal A}^{13}_3:}\quad {\mathfrak T}_1(t,x,u) =
(t,u,x),\quad {\mathfrak T}_2(t,x,u) = (t,u,\ln{u}),\quad q = 2,\\
&&{\mathcal A}^{14}_3:\quad {\mathfrak T}_1(t,x,u) =
(t,x,u),\quad {\mathfrak T}_2(t,x,u) = (t,x^{q-2}u^{q-1},
\frac{1}{q-2}\ln{u}),\quad q \ne 2,\\
&&\phantom{{\mathcal A}^{14}_3:}\quad {\mathfrak T}_1(t,x,u) =
(t,x,u),\quad {\mathfrak T}_2(t,x,u) = (t,u,\ln{u}),\quad q = 2,\\
\end{eqnarray*}

Now if an equation invariant under one of the algebras from the list
${\mathcal A}^1_3$ - ${\mathcal A}^6_3$, ${\mathcal A}^{10}_3$ -
${\mathcal A}^{14}_3$ admits additional Lie symmetry satisfying
(\ref{1.7.2}), then every member of the sequence of equations
obtained by successive application of transformations ${\mathfrak P}
\circ {\mathfrak T}_2 \circ {\mathfrak P} \circ {\mathfrak T}_1
\circ {\mathfrak P}$ possesses nonlocal symmetry. Moreover, if we
are able to construct an exact solution of some equation from the
sequence, we can map this solution into solutions of the remaining
equations from the sequence.

For the algebras ${\mathcal A}^7_3$ - ${\mathcal A}^9_3$, ${\mathcal
A}^{15}_3$ - ${\mathcal A}^{18}_3$, and ${\mathcal A}^{20}_3$ -
${\mathcal A}^{22}_3$ the algorithm stops after the second
iteration, since the third symmetry operator turns into a nonlocal
one. Below we give the corresponding sequences of transformations.
\begin{eqnarray*}
&&{\mathcal A}^7_3:\quad {\mathfrak P} \circ {\mathfrak P},\\
&&{\mathcal A}^8_3:\quad {\mathfrak P} \circ {\mathfrak T}_1
\circ {\mathfrak P},\\
&&{\mathcal A}^9_3:\quad {\mathfrak P} \circ {\mathfrak P},\\
&&{\mathcal A}^{15}_3:\quad {\mathfrak P} \circ {\mathfrak T}_1
\circ {\mathfrak P},\\
&&{\mathcal A}^{16}_3:\quad {\mathfrak P} \circ {\mathfrak P},\\
&&{\mathcal A}^{17}_3:\quad {\mathfrak P} \circ {\mathfrak T}_1
\circ {\mathfrak P},\\
&&{\mathcal A}^{18}_3:\quad {\mathfrak P} \circ {\mathfrak P},\\
&&{\mathcal A}^{20}_3:\quad {\mathfrak P} \circ {\mathfrak T}_2
\circ {\mathfrak P},\\
&&{\mathcal A}^{21}_3:\quad {\mathfrak P} \circ {\mathfrak T}_2
\circ {\mathfrak P},\\
&&{\mathcal A}^{22}_3:\quad {\mathfrak P} \circ {\mathfrak T}_2
\circ {\mathfrak P},
\end{eqnarray*}
where
$$
{\mathfrak T}_1(t,x,u) = (t,u,x),\quad
{\mathfrak T}_2(t,x,u) = (t,x^3u,\frac{1}{3}\ln{u}).
$$
Applying the above sequences of transformations to the corresponding
invariant equations yield evolution equations of the form
(\ref{0.1}) that admit nonlocal symmetries. The explicit form of
those symmetries is very cumbersome, therefore we present the
nonlocal symmetries obtained through the action of ${\mathfrak P}$
only. In addition, we give the corresponding transformed equations.
\begin{eqnarray}
&&{\mathcal A}^7_3:\quad e_3 = x^2\partial_x + (v -
xu)\partial_u,\quad
u_t = \frac{\partial}{\partial x}\Bigl(xf(t,x^3u_x)\Bigr),\nn\\
&&{\mathcal A}^8_3:\quad e_3 = (x + v)\partial_x - u\partial_u,\quad
u_t = \frac{\partial}{\partial x}\Bigl(u\exp(u^{-1})
f(t,u^{-3}\exp(u^{-1})u_x\Bigr),\nn\\
&&{\mathcal A}^9_3:\quad e_3 = x^2\partial_x +((1 - x)u + v)\partial_u,\quad
u_t = \frac{\partial}{\partial x}\Bigl(x\exp(-x^{-1})
f(t,x\exp(x^{-1})u_x)\Bigr),\nn\\
&&{\mathcal A}^{15}_3:\quad e_3 = v\partial_x - (u + 1)\partial_u,\quad
u_t = \frac{\partial}{\partial x}\Bigl((1 - u^2)^{1/2}
f(t,(1 + u^2)^{-3/2}u_x)\Bigr)\nn\\
&&{\mathcal A}^{16}_3:\quad e_3 = (1 + x^2)\partial_x +
(v - xu)\partial_u,\quad u_t = \frac{\partial}{\partial x}
\Bigl((1 + x^2)^{1/2}f(t,(1 + x^2)^{3/2}u_x)\Bigl),\nn\\
&&{\mathcal A}^{17}_3:\quad e_3 = (qx + v)\partial_x -
(u + 1)\partial_u,\quad u_t = \frac{\partial}{\partial x}
\Bigl((1 + u^2)^{1/2}\exp(-q\arctan{u})\nn\\
&&\quad\times f(t,(1 + u^2)^{-3/2}\exp(-q\arctan{u})u_x)\Bigr),
\label{3.7}\\
&&{\mathcal A}^{18}_3:\quad (1 + x^2)\partial_x +
((q - x)u + v)\partial_u,\quad u_t = \frac{\partial}{\partial x}
\Bigl((1 + x^2)^{1/2}\exp(\arctan{x})\nn\\
&&\quad\times f(t,(1 + x^2)^{3/2}\exp(-q\arctan{x})u_x)\Bigr),\nn\\
&&{\mathcal A}^{20}_3:\quad e_2 = xv\partial_x -
(xu + 3uv)\partial_u,\quad u_t = \frac{\partial}{\partial x}
\Bigl(xuf(t,x^{-5}u^{-3}u_x + 2x^{-6}u^{-2})\Bigr),\nn\\
&&{\mathcal A}^{21}_3:\quad e_2 = xu\partial_x -
(4x^{-5} + xu + 3uv)\partial_u,\quad u_t =
\frac{\partial}{\partial x}\Biggl(x^{-2}(4 + x^6u^2)^{1/2}\nn\\
&&\quad\times f\Bigl(t,(4 + x^6u^2)^{-3/2}
(x^4u_x + 5x^3u + \frac{1}{2}x^9u^3)\Bigr)\Biggr),\nn\\
&&{\mathcal A}^{22}_3:\quad e_2 = xu\partial_x +
(4x^{-5} - xu - 3uv)\partial_u,\quad u_t =
\frac{\partial}{\partial x}\Biggl(x^{-2}(x^6u^2 - 4)^{1/2}\nn\\
&&\quad\times f\Bigl(t,(x^6u^2 - 4)^{-3/2}
(x^4u_x + 5x^3u - \frac{1}{2}x^9u^3)\Bigr)\Biggr).\nn
\end{eqnarray}
Here $v = \partial_x^{-1}u$ is the nonlocal variable.

Finally, when computing transformations associated with the algebra
${\mathcal A}^{19}_3$ we have to stop after the first iteration,
since the second and third basis operators are mapped by ${\mathfrak
P}$ into nonlocal symmetries
\begin{eqnarray}
&&e_2 = \cos{u}\partial_x + (u\sin{v} + \sec^2{x}\sin{v} +
u\cos{v}\tan{x})\partial_u,\nn\\
&&e_3 = -\sin{u}\partial_x + (u\cos{v} + \cos{v}\sec^2{x} -
u\sin{v}\tan{x})\partial_u,\label{3.8}
\end{eqnarray}
where $v = \partial_x^{-1}u$. So that the sequence of differential
transformations boils down to a single transformation ${\mathfrak T}
= {\mathfrak P}$. And what is more, the evolution equation invariant
under the nonlocal symmetries $e_2$ and $e_3$ reads as
$$
u_t = (\sec^2{x}\tan{x} + uu_x)(\sec^2 x + u^2)^{-1/2}f +
(\sec^2{x} + u^2)^{1/2}\frac{\partial f}{\partial x}
$$
with
$$
f = f\Bigl(t,(u_x\cos{x} - (2 + u^2\cos^2{x})u\sin{x})
(1 + u^2 \cos^2{x})^{-3/2}\Bigr).
$$

Let us emphasize that symmetries (\ref{3.7}), (\ref{3.8}) contain
nonlocal variable and, consequently, cannot be obtained within the
framework of the traditional infinitesimal Lie method.

\section{Concluding Remarks}

We develop the regular method for group classification of evolution
equations admitting nonlocal symmetries. The starting point of the
method is an evolution equation that possesses nontrivial Lie
symmetry. The source of nonlocal symmetries are classical symmetries
subjected to nonlocal transformations.

One of the by-products of our approach are the sequences of
evolution equations related to the initial invariant equation. So
that if we succeed in constructing a (general or particular)
solution of the sequence member then it is possible to map this
solution into solutions of the remaining equations from the
sequence.

Of special interest is the case when one of the members of the
sequence is the linear evolution equation, since this means that the
remaining equations are linearizable. Moreover, a direct application
of the method to linear evolution equations admitting nontrivial Lie
symmetries is also of importance, since it yields sequences of
linearizable evolution equations.

These and related problems are under study now and will be reported
elsewhere.

\end{document}